\pdfoutput=1
\documentclass[final,authoryear,5p]{elsarticle}
\usepackage{color}
\usepackage{tablefootnote}
\def\aj{AJ}%
\def\araa{ARA\&A}%
\def\apj{ApJ}%
\def\apjl{ApJ}%
\def\apjs{ApJS}%
\def\aap{A\&A}%
\def\mnras{MNRAS}%
\def\nar{New A Rev.}%
\def\pasp{PASP}%
\def\pasj{PASJ}%
\usepackage{graphics}
 \usepackage{graphicx}
\usepackage{epsfig}
\usepackage{subfig}
\usepackage{amssymb}
\usepackage[a4paper=true,%
breaklinks=true,%
colorlinks=true,%
pdfauthor={Tremou et al.},%
pdftitle={Demonstration of KVN Phase Referencing Capability}%
]{hyperref}

\journal{Advances in Space Research}

\begin{document}

%%%%%%%%%%%%%%%%%%%%%%%%%%%%%%%%%%%%%%%%%%%%%%%%%%%%%%%%%%%%%%%%%%%%%%%%%%%%%
%% Frontmatter
\begin{frontmatter}


% Use the tnoteref command within \title and fnref within \author or \address for footnotes;
% use the corref command within \author for corresponding author footnotes;
% \author{Name\corref{cor1}\fnref{label2}}
% \address{Address\fnref{label3}}

\title{Demonstration of KVN Phase Referencing Capability}
%\tnotetext[footnote1]{This template can be used for all publications in Advances in Space Research.}

% Use optional labels to link authors explicitly to addresses:

\author[label1,label2]{Evangelia Tremou}\corref{cor1}
%\address{Yonsei University Observatory, Seoul 120-749, Republic of Korea}
%\fntext[footnote2]{Additional information regarding the corresponding author}
\ead{tremou@msu.edu}

% Url can be given like this:
% \ead[url]{http://www.elsevier.com/wps/find/authorsview.authors/latex}

\author[label3]{Taehyun Jung}
%\address{Korea Astronomy and Space Science Institute, 776 Daedeokdae-ro, Yuseong-gu, Daejeon, 305-348, \ Republic of Korea}
%\fntext[footnote3]{Additional information about the second and third authors}
\ead{thjung@kasi.re.kr}

\author[label1]{Aeree Chung}
%\address{Department of Astronomy and Yonsei University Observatory, Yonsei University, 120-749, Seoul, \ Republic of Korea}
%\fntext[footnote4]{Additional information about the co-authors}
\ead{achung@yonsei.ac.kr}

\author[label3]{Bong Won Sohn}
%\address{Korea Astronomy and Space Science Institute, 776 Daedeokdae-ro, Yuseong-gu, Daejeon, 305-348, \ Republic of Korea}
%\fntext[footnote4]{Additional information about the co-authors}
\ead{bwsohn@kasi.re.kr}

\cortext[cor1]{Corresponding author}

\address[label1]{Department of Astronomy and Yonsei University Observatory, Yonsei University, 120-749, Seoul, \ Republic of Korea}
\address[label2]{Department of Physics and Astronomy, Michigan State University, East Lansing, Michigan 48824, \ USA}
\address[label3]{Korea Astronomy and Space Science Institute, 776 Daedeokdae-ro, Yuseong-gu, Daejeon, 305-348, \ Republic of Korea}

\begin{abstract}

We present the results of Very Long Baseline Interferometry (VLBI) observations using the phase reference technique to detect weak Active Galactic Nuclei (AGN) cores in the Virgo cluster. Our observations were carried out using the Korean VLBI Network (KVN).
We have selected eight representative radio galaxies, seven Virgo cluster members and one galaxy (NGC 4261) that is likely to be in the background. The selected galaxies are located in a range of density regions showing various morphology in 1.4 GHz continuum. 
Since half of our targets are too weak to be detected at K-band we applied a phase referencing technique to extend the source integration time by calibrating atmospheric phase fluctuations.
We discuss the results of the phase referencing method at high frequency observations and we compare them with self-calibration on the relatively bright AGNs, such as M87, M84 and NGC 4261.
In this manuscript we present the radio intensity maps at 22 GHz of the Virgo cluster sample while we demonstrate for first time the capability of KVN phase referencing technique.

\end{abstract}


\end{frontmatter}

\parindent=0.5 cm

%%%%%%%%%%%%%%%%%%%%%%%%%%%%%%%%%%%%%%%%%%%%%%%%%%%%%%%%%%%%%%%%%%%%%%%%%%%%%
%% Main text
\section{Introduction}
Astronomical observations at short wavelength suffer from atmospheric disturbances. 
Especially, high frequency (in millimeter wavelength) observations using radio interferometers such as Atacama Large Millimeter/   submillimeter Array (ALMA) \footnote{http://www.almaobservatory.org/} and Very Long Baseline Array (VLBA)\footnote{http://www.vlba.nrao.edu/} are severely affected by atmospheric conditions.
The electromagnetic waves pass through the troposphere where the water vapour is present before they reach the interferometer.  The irregular  distribution of the water vapour causes phase fluctuations and delays in the coherency between the signals received by each interferometric element. 
Therefore, in order to improve the accuracy of observations, the atmospheric phase fluctuations should be corrected. 

In principle, those atmospheric effects can be reduced by two main approaches.  
One way would be the direct measurement of the integrated water vapour content by the water vapour radiometer \citep{1981RaSc...16..235M, 1998AAS...192.8103M, 2004evn..conf..265R}. 
These measurements require almost perfect weather conditions, which is hardly the case and the detection of the atmospheric phase fluctuations is limited above several tens of seconds. 
Instead, the phase referencing (PR) technique \citep{1979AJ.....84.1459S, 1988IAUS..129..523A}  turns out to be very useful to correct the phases of the target using the phases of a strong calibrator. With the PR method the image sensitivity and the dynamic range can be increased and low detection thresholds can potentially be achieved up to the sub-mJy level at low frequencies \citep[$<$5 GHz;][]{1995ASPC...82..327B}. A strong calibrator needs to be observed frequently by rapidly switching the antenna between a calibrator and a target in order to calibrate the visibilities of the scientific target.

The PR can be applied through several different ways depending on the instrument. Some of the methods that have been used to perform phase referencing are: 
fast antenna switching \citep{1979AJ.....84.1459S, 1995ASPC...82..327B, 1996PASP..108..520W}, paired antenna methods \citep{1996RaSc...31.1615A, 1998RaSc...33.1297A}, dual beam antennas \citep[e.g. in VERA;][]{2003PASJ...55L..57H}, fast frequency switching  \citep[e.g. in VLBA;][]{2005A&A...433..897M}, cluster-cluster mode \citep{2002evn..conf...57R, 2003ASPC..306...39P}, and  multi-frequency feed \citep{2003ASPC..306...53S, 2011PASJ...63..717J}.

The quality as well as the success of various PR techniques listed above are limited by residual errors in the differential phases between a pair of sources (target and calibrator). The observations of two sources result in a differential excess path length that can seriously affect  the scientific result \citep{1995ASPC...82..327B}. Therefore the phase reference requires a calibrator very close to a target, which is not always the case in practice. In order to get a better handle on this issue, bigradient phase referencing (BPR) has been suggested by \cite{2006PASJ...58..777D}, which utilizes nearby weak sources as calibrators of a target.

In this article, we present the results of Korean VLBI Network (KVN)\footnote{http://kvn.kasi.re.kr} PR performance at K-band (22 GHz). 
The sample and its selection criteria are described in detail in Section \ref{sample}. Our observation strategy and the data reduction are discussed in Section \ref{data}. 
In Section \ref{results}, we comment on the primary outcome of this work.
Finally, we summarize and discuss our results in Section \ref{sum}.

\section{The sample}\label{sample}

We have selected a representative sample of radio bright galaxies in the Virgo cluster. Virgo, as the nearest rich galaxy cluster \citep[$\sim$16.5 Mpc,][]{2007ApJ...655..144M}, is an ideal place to study detailed properties of individual galaxies. Also it is dynamically young and hence contains a range of density environments, providing an ideal laboratory to study how galactic properties change with the surroundings. The two main goals for this study are a) to test KVN PR feasibility and b) to investigate the AGN activities in different density environments. While more scientific results will be presented in a follow-up paper, we focus on the KVN PR feasibility in this work. 

\begin{figure}[t!]
\begin{center}
\includegraphics*[width=9cm,,natwidth=610,natheight=642]{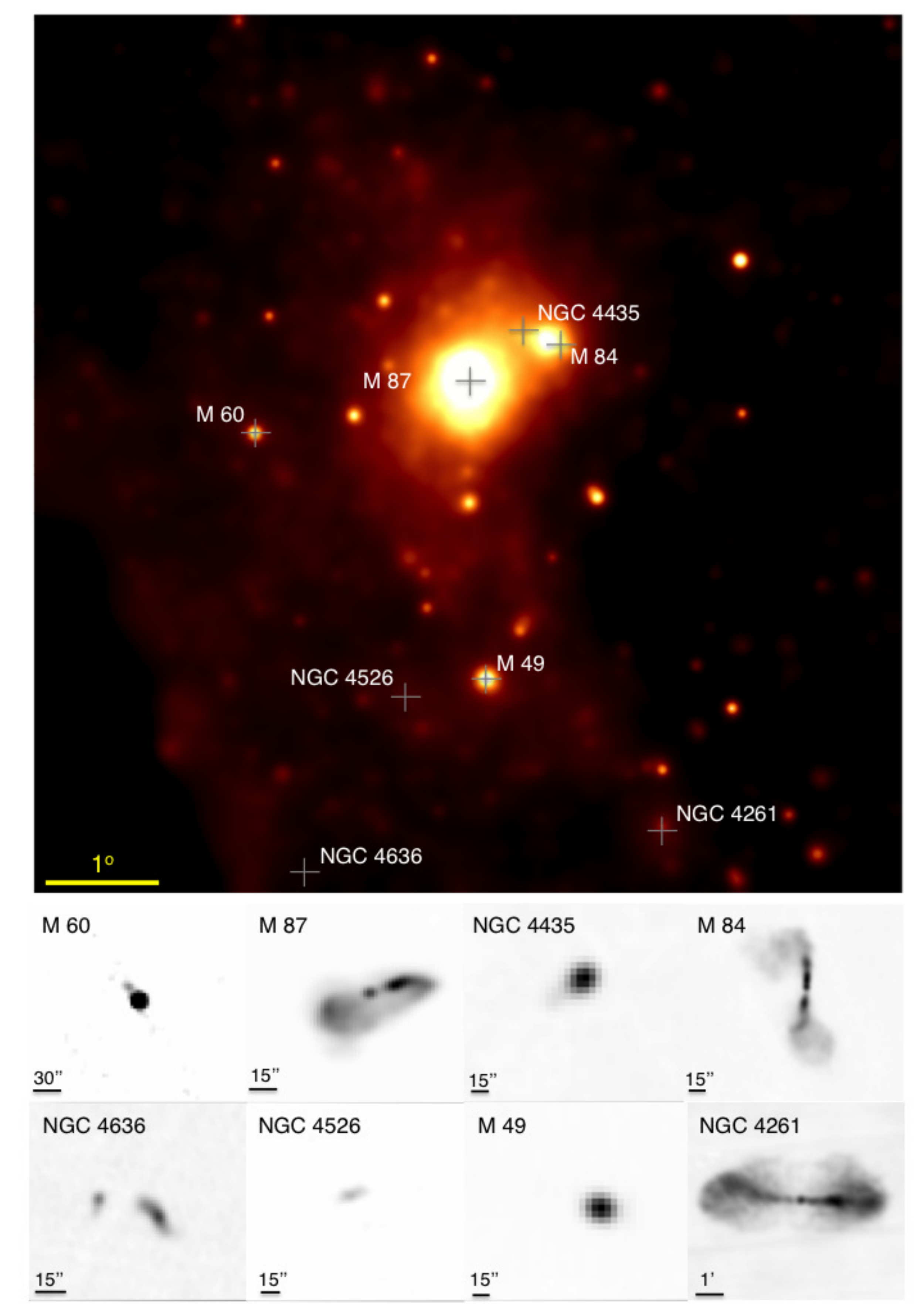}
\end{center}
\vspace{-0.5cm}
\caption{In the top, the locations of the sample in the ROSAT X-ray image of the Virgo cluster (orange background). The white bar on the bottom left represents 1 degree. The 1.4 GHz continuum images from the FIRST survey are shown in grey-scale on the bottom rows.}
\label{first_virgo}

\end{figure}

To achieve our goals, we have selected 8 sources, 7 Virgo members and one galaxy (NGC 4261) that is likely to be background source but still close enough to be studied in high resolution. The galaxies in the sample are either radio bright \citep{2009AJ....138.1990C,2005A&A...435..521N} with extended radio features and/or optically identified LINER or Seyfert. The selected galaxies are located in a range of density regions showing various morphology in 1.4 GHz continuum (Figure \ref{first_virgo}) radio maps from the Very Large Array (VLA) Faint Images of the Radio Sky at Twenty-Centimeters (FIRST) survey \citep{1994ASPC...61..165B}. 

\textbf{NGC 4261} reveals quite symmetric kiloparsec-scale jets that are well aligned as shown in a VLBA imaging study by \cite{1997AAS...19110403J}. Its morphology and spectral index distribution indicate that free-free absorption is not significant within the central region but it affects the counter-jet. The radio emission decreases quickly in both directions with a large opening angle, indicating rapid expansion. This could occur when the internal pressure is lower than the surrounding medium. The comparison of high frequency radio observations at 22 GHz and 43 GHz with lower radio frequencies, shows a brightness asymmetry that can be caused by the presence of an edge on inner accretion disc  \citep{1999AAS...194.5018J}. Additionally, this galaxy reveals strong optical emission from the center, UV emission from jet and the nucleus shows short-term  X-ray variability \citep{2003ApJ...586L..37S}.

\textbf{M 84} shows a slightly extended core to the north at high frequency (43 GHz) VLBI observations by \cite{2004AJ....127..119L}.
The spectral energy distribution of its nucleus is comparable to BL Lac-like objects. The ionized gas in the center of the source seems to be associated with non stellar process instead of hot stars \citep{2000ApJ...534..189B}.
\cite{2008ApJ...686..911F} studied X-ray data in depth showing that the overpressure of the cavity (in respect to the medium) resulted in energetic waves, which have a substantial contribution to the total energy of AGN. \cite{2010ApJS..187..135E} classified M84 as low-luminosity AGN based on weak optical lines, which could be dust-obscured.

\textbf{NGC 4435} shows a strong core with a short radio tail. It is close to NGC 4438 on the sky while it is questionable whether they are physically associated and hence this galaxy also has gone through some tidal interactions or not \citep[e.g.][]{2005astro.ph..7252V}.

\textbf{M 49} is  a luminous LINER galaxy and has a  kinematically distinct core \citep{1988ApJS...68..409D}. At 6 cm it shows a linearly polarized emission which may be associated with outflow and jet-like features.

\begin{table}[t!]\footnotesize
\caption{General properties of the sample.}
	\centering
		\begin{tabular}[width=0.4\textwidth]{ccccr}
		\hline
RA  & DEC 		& Target & Velocity & Activity\\
(J2000) & (J2000)& 		& (km/s) &Type\\
\hline
 12:19:23.22 &+05:49:30.8 &NGC 4261 & 2212 & LINER \\

 12:25:03.74 &+12:53:13.1	&M84 & 1017 &Seyfert 2\\

12:27:40.49 & +13:04:44.2	&NGC 4435 & 791 &\\

12:29:46.76 & +08:00:01.7 &M49 & 981 &Seyfert 2\\

 12:30:49.42 &+12:23:28.0 &	M87 & 1284  & LINER \\
 
12:34:03.09 & +07:41:58.3	&NGC 4526 & 617 & \\

12:42:49.83 & +02:41:16.0	&NGC 4636 & 938 & LINER \\
   	 		
12:43:39.98 & +11:33:09.7 & M60 & 1110 &\\
           	 		
		\hline
		\end{tabular}
	
\label{tab:sample}
\vspace{-0.3cm}
\end{table}

\textbf{M 87} is the most famous and well studied galaxy among our sample and it has already been observed in a wide range of radio frequencies. The galaxy shows ionized gas filaments, but no  [O\,{\sc{iii}}] emission has been detected along the radio jet, which might be associated with shock excitation phenomena. 

\textbf{NGC 4526} is a  core S{\'e}rsic galaxy \citep{2009AJ....138.1990C} implying the presence of a flat central core. This galaxy is also known for its relatively high FIR/radio ratio, for which ram pressure is likely to be responsible for \citep{2009ApJ...694.1435M}. 

\textbf{NGC 4636} shows an asymmetric morphology and its bright jets flowing into fainter extended ($\sim$ 2.5 kpc) radio lobes \citep{2011ApJ...732...95G}.  This galaxy has not been observed so far in higher frequencies than 15 GHz. It has spiral arm-like structures which are coincident with the border of two X-ray cavities. The north-east cavity is dominated with low frequency radio-emitting plasma, but in south-west side there is no activity detected in radio regime.

\begin{table*}[t!]\footnotesize
\caption{Parameters and Results of KVN Phase Referencing Observation at 22 GHz.}
	\centering
	\begin{minipage}{\textwidth}
	\renewcommand{\thefootnote}{\thempfootnote}
		\begin{tabular}[width=\textwidth]{ccccccc}
		\hline
		Target \footnote{Column 1 shows the target sample, column 2 shows the calibrator for each target and the column 3 shows the separation angle between a target and a calibrator. The on source time of the target is presented in the column 4 and the sensitivity at 10 $\sigma$ in the  column 5 according to the calculations from the European VLBI Network calculator (http://www.evlbi.org/cgi-bin/EVNcalc).The sensitivity is a function of the system equivalent flux density (SEFD$\sim$1300Jy), bandwidth and integration time. The expected fluxes have been estimated by extrapolating the archival data at 4.8 and 8.4 GHz \citep[][2$\sim$5 mas resolution, comparable to this work]{2009AJ....138.1990C,2005A&A...435..521N} are presented in column 6. In the column 7 we present the total flux density as a result of this work with KVN PR at 22 GHz.}& Calibrator\footnotemark[\value{mpfootnote}] & Separation angle\footnotemark[\value{mpfootnote}] &Target  on Source\footnotemark[\value{mpfootnote}] &  Image Sensitivity (10 $\sigma$)\footnotemark[\value{mpfootnote}] & Estimated Flux \footnotemark[\value{mpfootnote}]&Measured Flux \footnotemark[\value{mpfootnote}]\\
		& & (degrees) & Time (hours) & (mJy/beam) &Density(mJy)& Density(mJy) \\
			\hline
					  NGC 4261 & 3C273 &4.4843  & 0.558 &10.48&280 &279 \\
					  
			M84 & M87 & 1.4902 &0.615 & 9.97 & 289 &100 \\
			
			NGC 4435 & M87 &1.0307  & 0.558 & 10.48 & $<$1&no detection~\footnote{The image sensitivity of the non detected sources might be lower than the estimated at 10$\sigma$.}\\

				M49& M87 & 4.3975   &0.558 & 10.48& 27&no detection\footnotemark[\value{mpfootnote}]\\
				
		  M87 & M84 &1.4902  & 2.633 &4.82&2000 &1616 \\

			NGC 4526 & M87 &4.7578  &0.558 & 10.48 & 4 &no detection\footnotemark[\value{mpfootnote}]\\

			NGC 4636 & 3C273 &3.4870  &0.558 & 10.48 & 60 &no detection\footnotemark[\value{mpfootnote}]\\
				
			M60& M87&  3.2505  &0.558 & 10.48& 45 &no detection\footnotemark[\value{mpfootnote}]\\
			
				\hline
		\end{tabular}
	
\label{tab:separangle}
	\end{minipage}
\end{table*}

\textbf{M 60} is treated as a non-AGN source but it has different structure from other early type galaxies.
Studies on the morphology of the central regions in galaxies \citep{2002A&A...387..441X} have shown that early-type, non-AGN galaxies have a smooth distribution of light in their central 100 pc region. The smooth structure with no particular light deviation features the lack of significant amount of material in the nucleus (parsec scale). However, M60 does show a central structure, indicating the presence of an active nucleus. 

In Table \ref{tab:sample}, the general properties and the AGN type of our sample are summarized.

\section{Observations and data reduction}\label{data}

\subsection{Observations}

The observations were carried out on 7th and 8th of December 2012, using the KVN \citep{2004evn..conf..281K, 2011PASP..123.1398L}. KVN is comprised of three identical telescopes (KU:Ulsan telescope, KY:Yonsei telescope and KT:Tamna telescope), 21m in diameter each. The maximum baseline length is about 480 km in the north-south direction (KY-KT baseline). KVN is capable of recording data in four frequencies simultaneously, 22, 43, 86 and 129 GHz with the highest angular resolution of $\sim$ 1 mas at 129 GHz.
%Comparing to other radio interferometric telescopes such as VLBA and Expanded Very Large Array (EVLA), the KVN resolution can be placed in between 
 
We have selected an observing frequency of 22 GHz with the total bandwidth of 256 MHz (16 channels x 16 MHz per channel) for each consecutive observing epoch, resulting in $\sim$ 5 mas resolution. It is also interesting to note that the KVN resolution in K-band is intermediate between what can be reached by the the VLBA ($\sim$ 0.35 mas) and one of the most powerful connected arrays such as VLA A-configuration can reach ($\sim$ 89 mas) in K-band.
We applied PR technique for compensating the atmospheric phase fluctuations. The PR requires fast antenna switching between target and calibrator sources. 
 The rapid slewing speed of KVN antennas ( $\sim$ 3 degrees/sec) allowed us to minimize the source switching time between the target and the calibrator and thus good calibration results can be achieved.

As we described in Section \ref{sample}, we observed eight galaxies (M49, M60, M84, M87, NGC 4435, NGC 4526, NGC 4636, NGC 4261)  and two flux calibrators 3C286 and 3C273. For phase calibrators we used 3C273, M87 and M84 which are strong enough to be detected by themselves. 
We conducted PR observations by pairing up the sources as listed in Table \ref{tab:separangle}. The integration of each scan was set to be $\sim$ 30 seconds per source, having a switching cycle of one minute for phase referencing. 
The separation angles between calibrator and target varies per pair as listed in Table \ref{tab:separangle}. For each galaxy, the calibrator with the shortest possible separation angle was chosen to be used for phase referencing. Our samples having various separation angles (\textbf{$\sim1-4.8$ }degrees) are good candidates for proving the PR capability of the KVN. 

The data are correlated with DiFX software correlator \citep{Deller}. The time resolution of the correlator output has been set to be $\sim$ 1 second.

\subsection{Data Reduction}

The data reduction process is carried out using the Astronomical Image Processing System (AIPS) \citep{1981NRAON...3....3F} for the steps of phase and amplitude calibration while the Difmap software \citep{1997ASPC..125...77S} is used to produce the radio images. 

The correlated data are loaded in AIPS using {\tt FITLD}. Then, they are sorted in the order of time-baseline using {\tt MSORT} and an index table is created with {\tt INDXR} before calibration. The amplitude in cross-correlation spectra is corrected with {\tt ACCOR} for the errors in sampler thresholds. The tasks {\tt ANTAB} and {\tt APCAL} are used to calculate the amplitude gain solutions using the system temperature and the antenna gain curve information.

  \begin{figure*}[t!]
	\centering	
	\subfloat[]{\includegraphics*[width=0.5\textwidth,,natwidth=610,natheight=642]{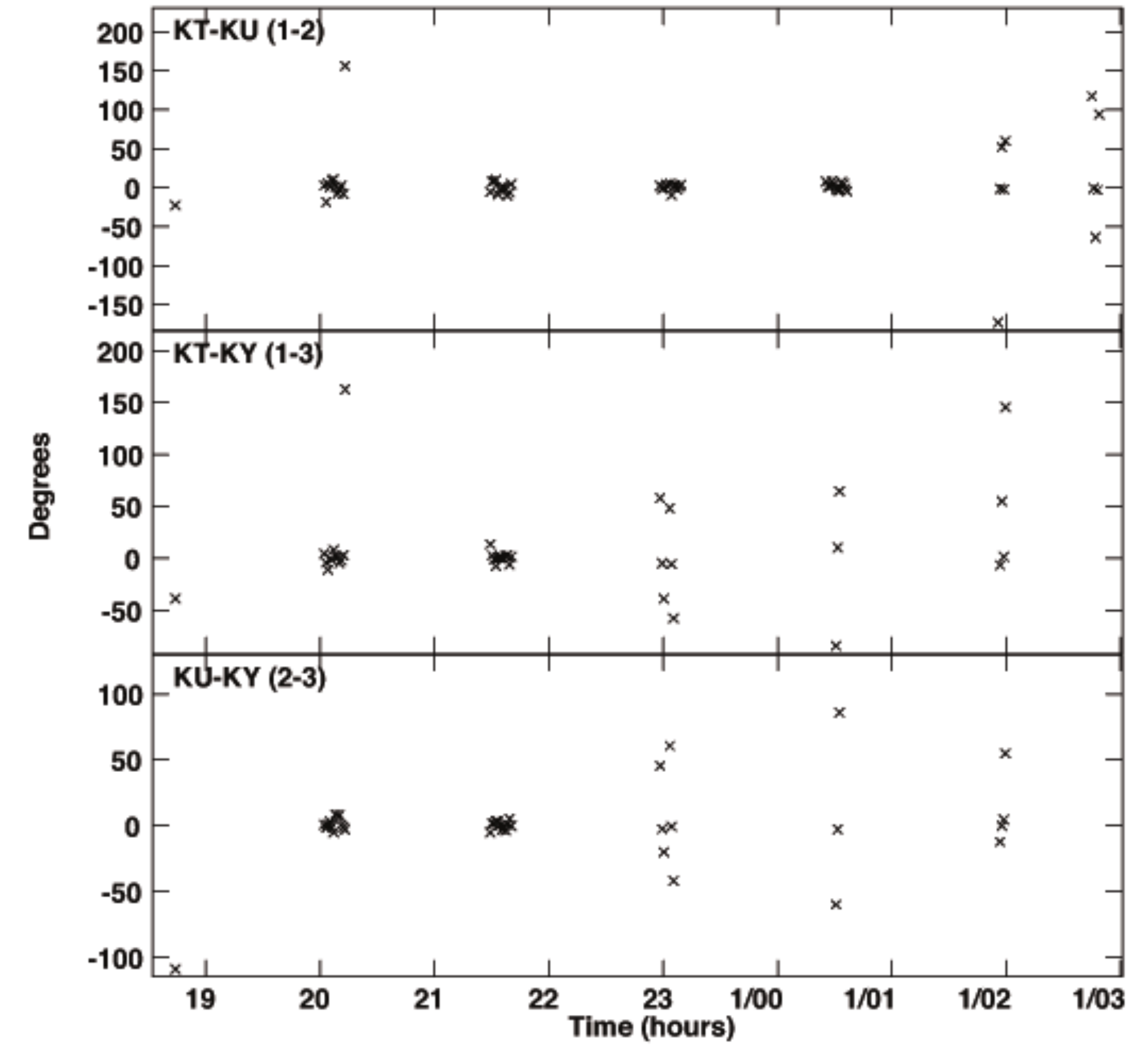}}
	\subfloat[]{\includegraphics*[width=0.5\textwidth,,natwidth=610,natheight=642]{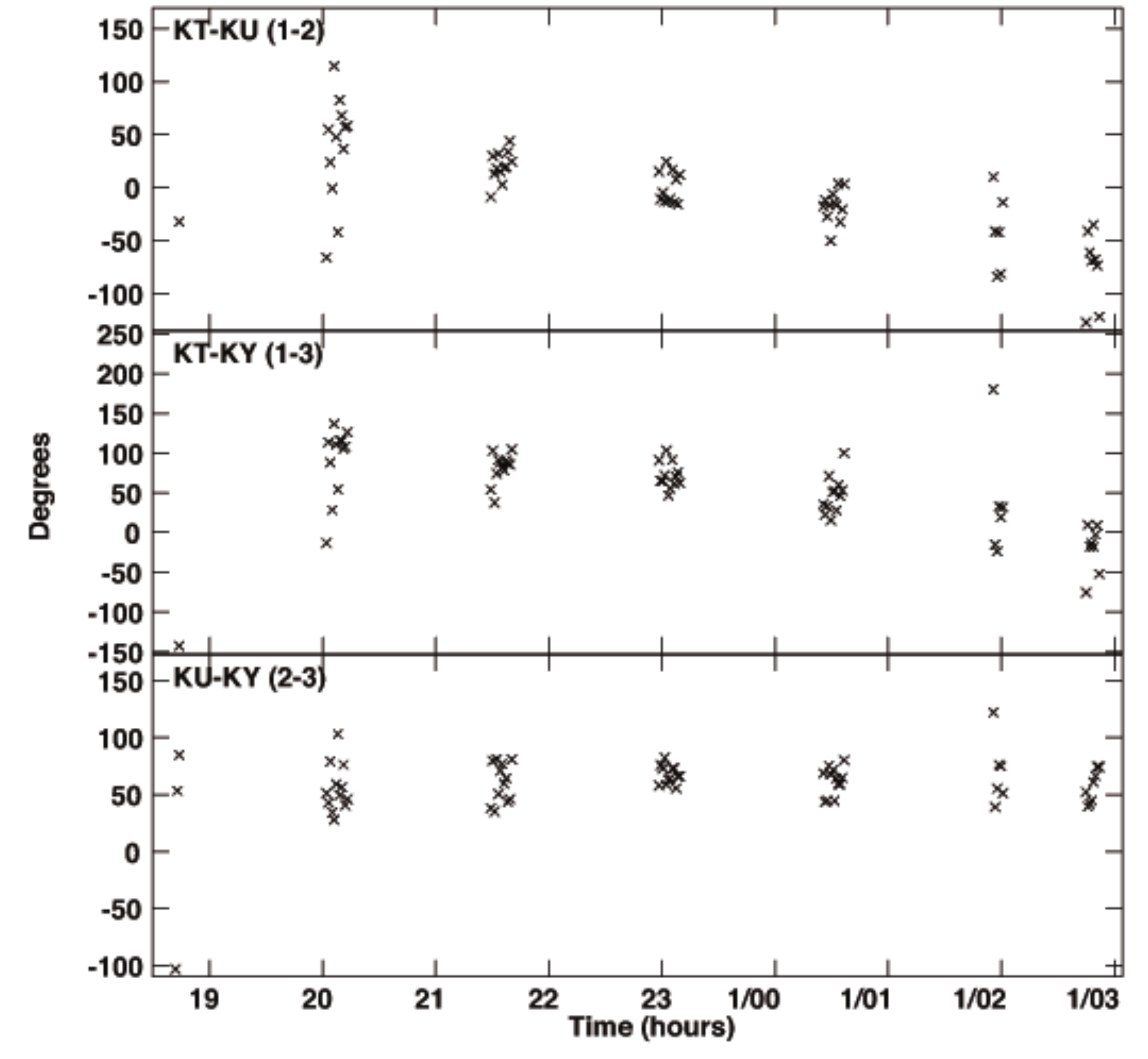}}
	\caption[The visibility phases after the phase calibration of M84 are presented for each baseline (KU stands for Ulsan telescope, KY, for Yonsei telescope and KT for Tamna telescope) after (a) applying self-calibration and after (b) applying the phase correction from the phase calibrator M87 (PR method).]
{The visibility phases after the phase calibration of M84 are presented for each baseline (KU stands for Ulsan telescope, KY, for Yonsei telescope and KT for Tamna telescope) after (a) applying a self-calibration and after (b) applying a PR from the phase calibrator M87.} 
\label{phasek12342am84}
\end{figure*} 

In VLBI observations, the path difference of interference fringes is not known until the differences among individual elements such as the clock of each station and the baseline geometry. The task {\tt FRING} allows us to calibrate out these effect by calculating delay residuals and the consequent time derivatives.

The three phase calibrators (M87, 3C273 and M84) have been used for our observations. %In order to find the most suitable calibrator we run task {\tt FRING} using either one of the three calibrators or all of them (multi-calibrators).
To create a calibration table, we have applied a linear vector interpolation by running {\tt CLCAL}. Then we run {\tt FRING} again by applying fringes of all calibrators to each source pair, and we apply the first degree of phase referencing. Using this outcome we have created the calibrated tables for each source group. 

In addition, we have tried phase-self calibration for three sources that could be detected without PR (M87, M84 and NGC 4261) to compare with the results from PR (Figure \ref{phasek12342am84}). 
 
Finally, all spectral channels in IF (Intermediate Frequency) were averaged by {\tt SPLIT}. After flagging the bad data, the {\tt CLEAN} algorithm and self-phase calibration were applied for obtaining the final radio maps  in Difmap.  

In this study we present  the phase solutions of  the first observation day since there is no significant difference in the data from the second day. The radio maps presented here have been produced from the combined data from both days.

\section{Results}\label{results}
\subsection{Phase Referencing and Imaging}

 \begin{figure}[t!]
	\centering	
	\includegraphics*[width=9cm,natwidth=610,natheight=642]{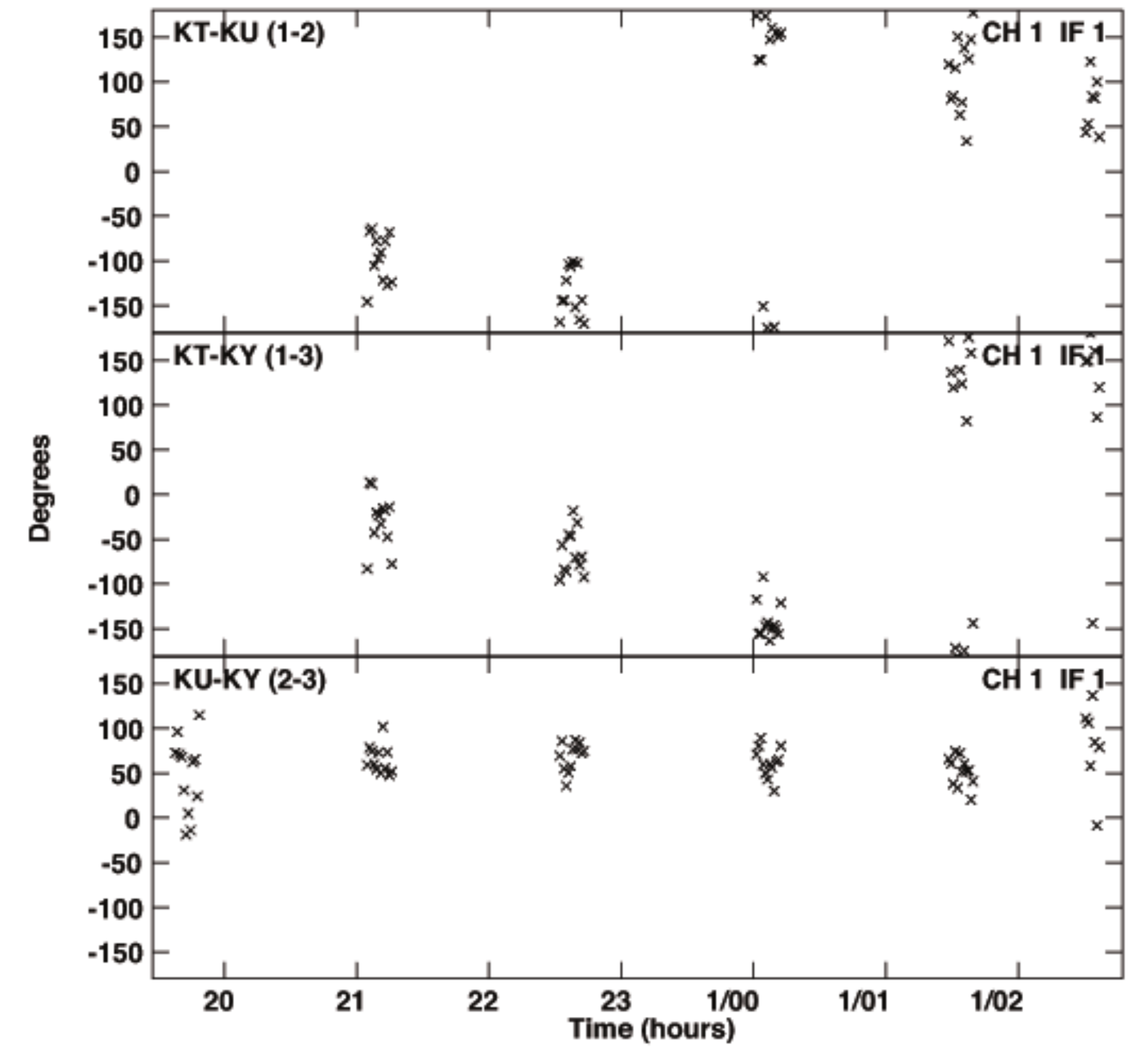}
	\caption[The visibility phases after the phase calibration for NGC 4261 are presented for each baseline.]
{The visibility phases after the phase calibration for NGC 4261 are presented for each baseline (KU stands for Ulsan telescope, KY, for Yonsei telescope and KT for Tamna telescope).} 
\label{phasek12342a}
\end{figure}

  \begin{figure*}[t!]
	\centering	
	\subfloat[]{\includegraphics*[width=0.3\textwidth,natwidth=610,natheight=642]{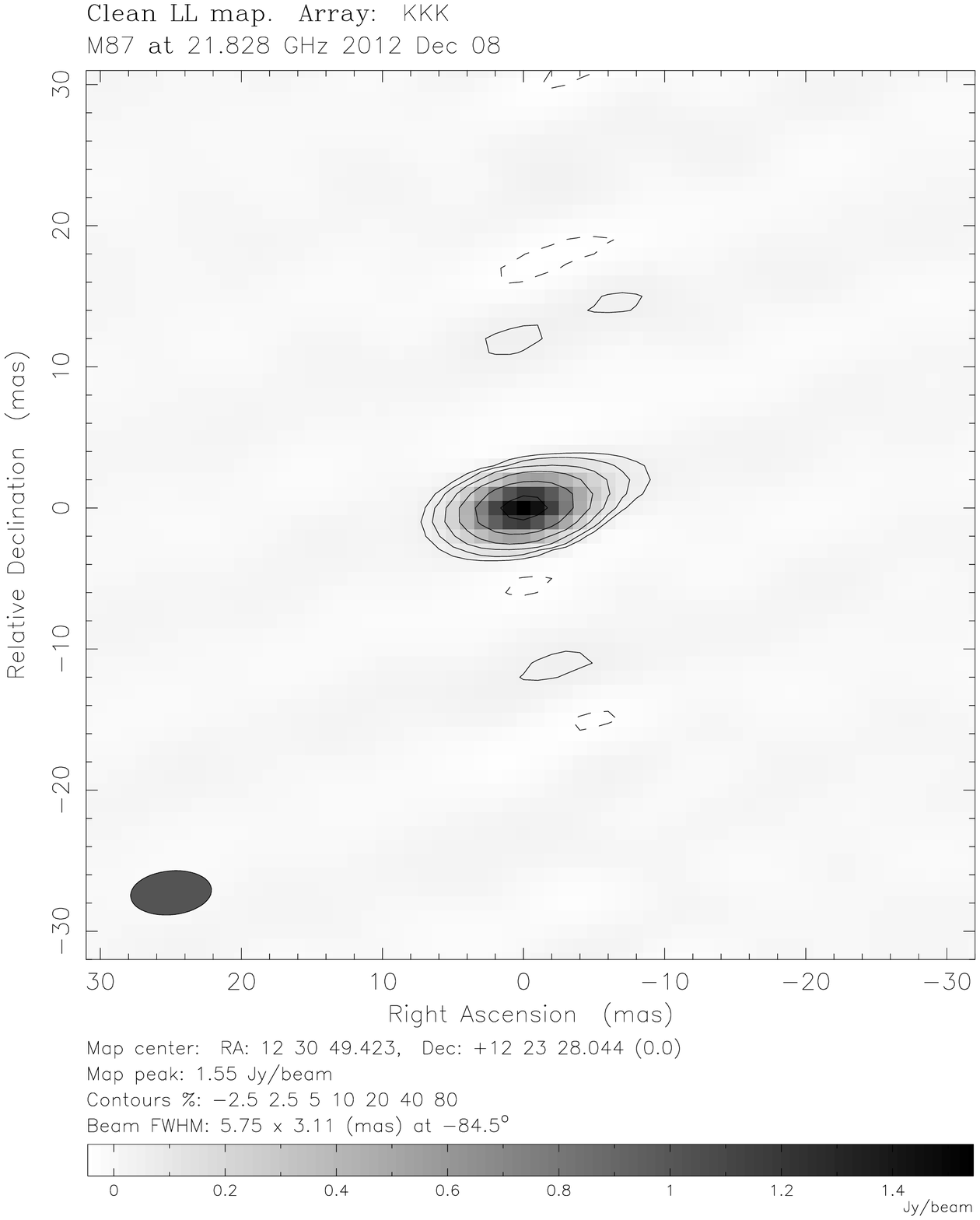}}
	\subfloat[]{\includegraphics*[width=0.3\textwidth,natwidth=610,natheight=642]{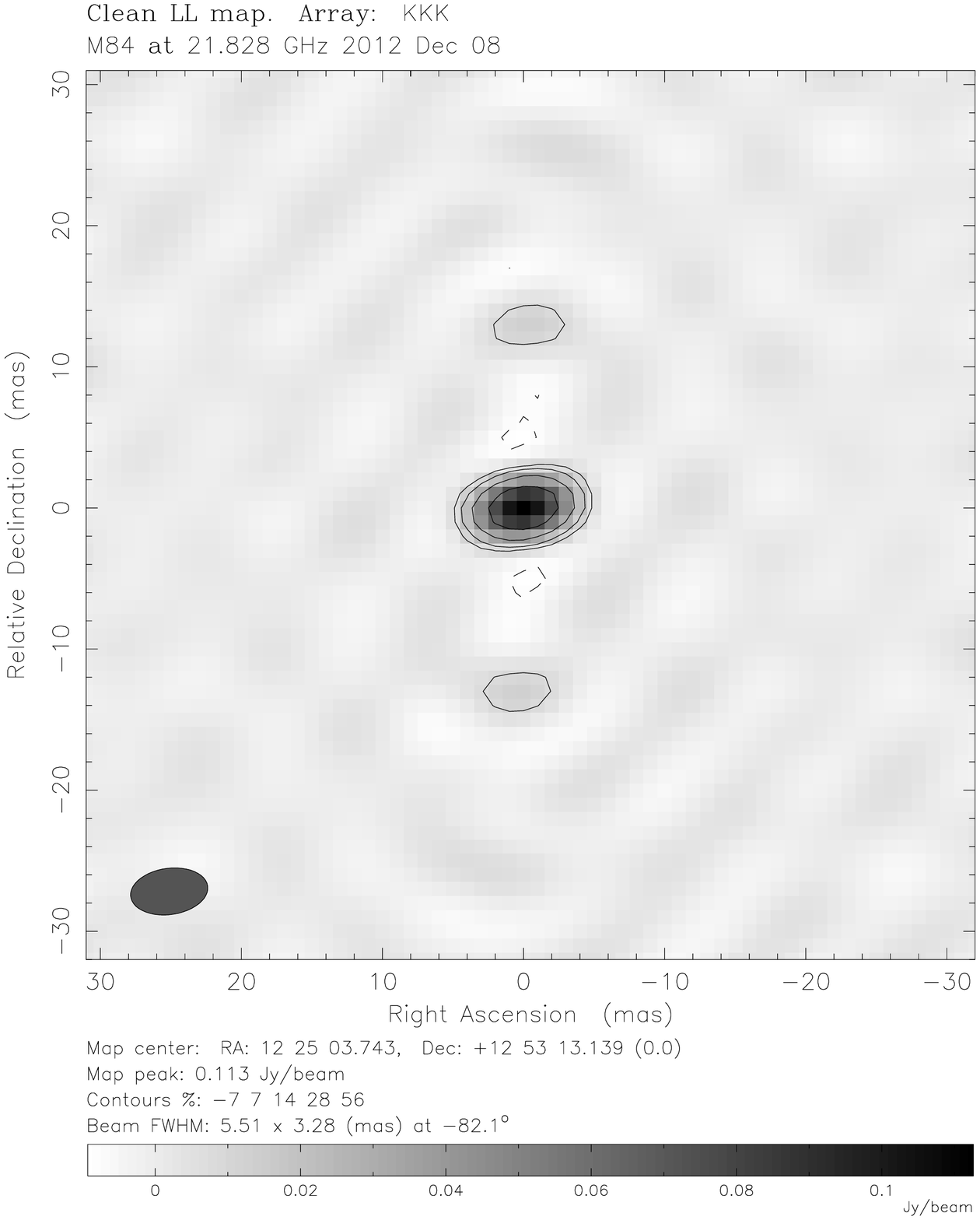}}
	\subfloat[]{\includegraphics*[width=0.3\textwidth,natwidth=610,natheight=642]{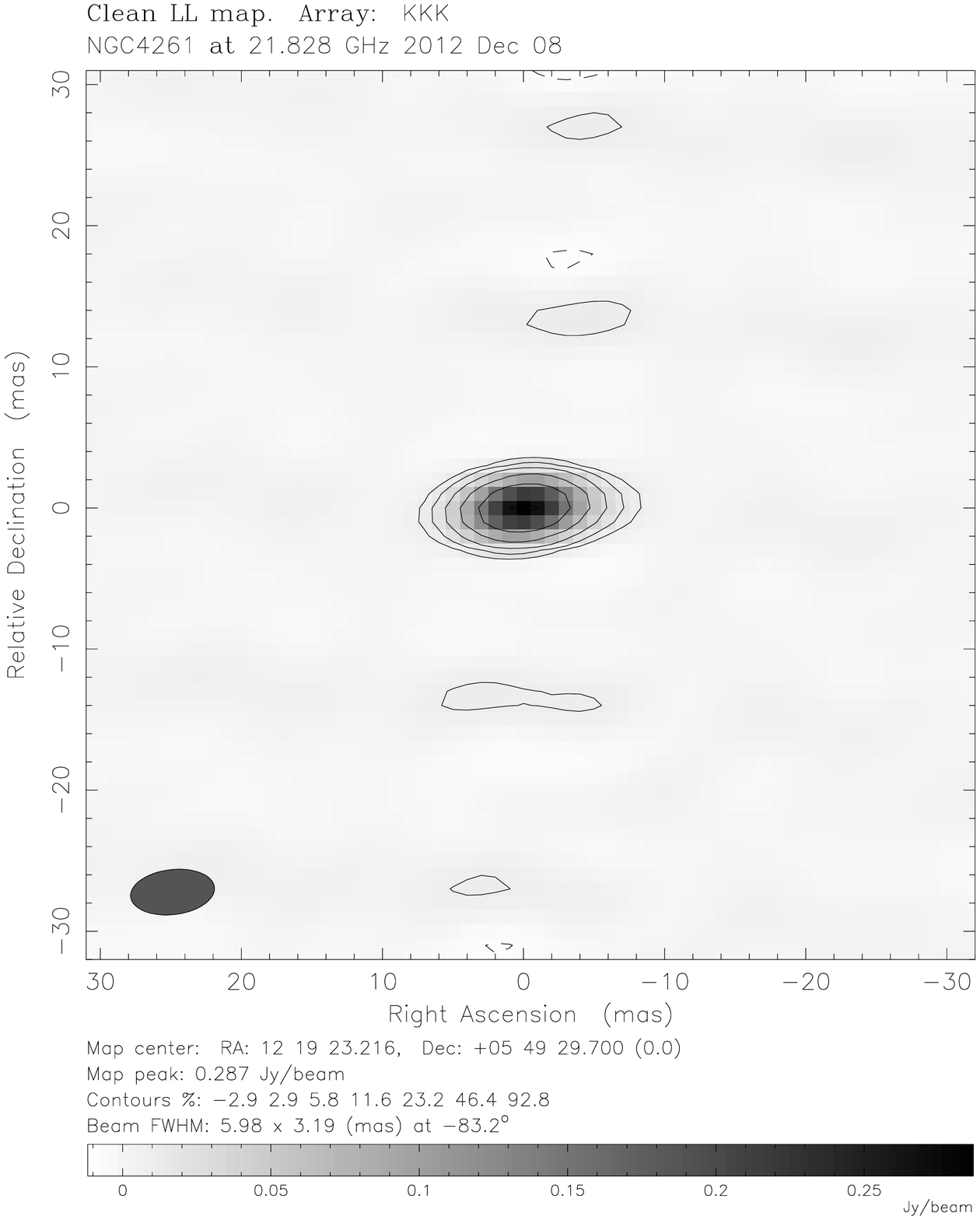}}
	\caption[KVN radio maps. (a) M87, (b)M84, (c) NGC 4261]
{\label{maps} KVN intensity maps at 22 GHz. (a) M87, (b) M84, (c) NGC 4261. The synthesized beam size varies from $\sim$ 5 to $\sim$ 6 mas. The \textbf{marginally} resolved core emission of M87 and NGC 4261 shows an elongated structure to the direction of their known extended jet in lower radio frequencies.} 
\end{figure*}

Fundamentally, PR uses the solutions of fringe phase of the calibrator to compensate the atmospheric fluctuations for the target. These solutions are time interpolated to the target observations resulting in calibrated target visibilities. 
Since the phases of calibrator are applied to the target, the shorter the separation angle is the more consistent the phase correction can be done.

Our sample provides excellent candidates to prove the phase referencing capability of KVN, because the source pairs of target and calibrator have various separation angles (\textbf{$\sim1-4.8$} degrees) and their expected fluxes range from a few tens mJy to several Jy at K-band. The expected flux at 22 GHz was extrapolated from measurements at lower frequencies (e.g. 4.8 and 8.4 GHz) assuming a power law for spectral distribution (Table \ref{tab:separangle}, estimated flux). These flux estimates include some uncertainties due to the measurement errors from different observations and due to source flux variability.

The visibility phases from the fringe fitting, are shown for M84 (Figure \ref{phasek12342am84}b) and NGC 4261 (Figure \ref{phasek12342a}). In the case of M84 and NGC 4261, the phase correction with the calibrators of M87 and 3C273, respectively, have improved the solutions (converged visibility phases) and thus the targets are clearly detected.   
%In the cases, of non detected sources, the visibility phases are scattered. Therefore, in Table \ref{tab:separangle}, we present the measured 22 GHz fluxes of KVN,  in comparison to the estimated fluxes from lower frequency observations and the estimated image thermal noise (image sensitivity). 
%We note four targets with no detection with a dynamic range of those maps $<\sim$20, which can be explained within the frame of the extrapolation estimations uncertainties. 

%What is remarkable is the impact of the separation angle on the non detected sources, although their low expected flux densities. These targets have the largest separation angles from their calibrators among our samples, which makes their detection challenging.  In spite of a relatively large distance from the calibrator, NGC 4261 is clearly detected. 
NGC 4261 is a relatively bright source, however the slope of the phase solutions profile in Figure \ref{phasek12342a}a shows the remarkable impact of the large separation angle. The separation angle affects, also, the non detected sources, although their low expected flux densities. These targets have the largest separation angles from their calibrators among our samples, which makes their detection challenging.  In spite of a relatively large distance from the calibrator, NGC 4261 is clearly detected. 

To better evaluate the PR outcome, we compare it with the self-calibration results. As it is shown in Figure \ref{phasek12342am84}, the self-calibration of phase (Figure \ref{phasek12342am84}a) for the KT-KY and KU-KY baselines is reliable till $\sim$23 UT(hours) and till $\sim$2 UT(hours) for the KT-KU baseline, where the detectability becomes challenging due to the scatter in the visibility phases. The former case is a result of the weather conditions. The system temperature at Yonsei site changed dramatically after $\sim$23 UT.
On the other hand, PR stays more stable throughout the entire observing run. In both cases the low elevation of the target can explain the dispersion of the visibilities at the beginning ($\sim$19 UT) and at the end (2$\sim$3 UT) of the observation. 

In particular, the PR technique can be severely affected by the low elevation of both calibrator and target, which is illustrated in Figure \ref{phasek12342am84}b between 19 UT$\sim$20 UT, where both M87 and M84 have low elevation. Furthermore, the peak flux and the intensity map structure of M84 have shown no notable difference between either cases (PR and self-calibration).

The images of M87, M84 and NGC 4261 at 22 GHz by KVN are presented in Figure \ref{maps}. The rms noise levels are 8, 0.9, 2 mJy/beam for M87, M84 and NGC 4261, respectively. The beam size varies from $\sim$ 5 to $\sim$ 6 mas.    Considering a distance of $\sim$ 16.5 Mpc, M87 extends to $\sim$ 14 mas ($\sim$ 1.1 pc), M84 to $\sim$ 9 mas ($\sim$ 0.95 pc) and NGC 4261 to $\sim$ 12 mas ($\sim$ 0.7 pc).
M87 is the brightest galaxy with flux density of 1616 mJy among the detections. 
The emission of M87 (Figure \ref{maps}a) and NGC 4261 (Figure \ref{maps}c) shows an elongated structure to the direction of their known extended jet at lower radio frequencies.

\subsection{Coherence}

In order to check the practical coherence time  at 22 GHz with the KVN, we have tested the signal-to-noise (SNR) increment along the integration time. The SNR is proportional to the square root of the bandwidth and the integration time. The first scan of 3C273 with $\sim$ 10 minutes length was used for this analysis.

Coherence time is dependent on the observing frequency and  has been empirically determined for atmospheric conditions where the telescopes are located \citep[e.g. $> \sim$ 1hr at 8.4 GHz for the VLBA with a switching angle of 1.9 deg between a source and a target;][]{1995ASPC...82..327B}. 

\begin{figure}[h!]
\vspace{0.1cm}
\centering
\includegraphics*[width=0.5\textwidth,natwidth=610,natheight=642]{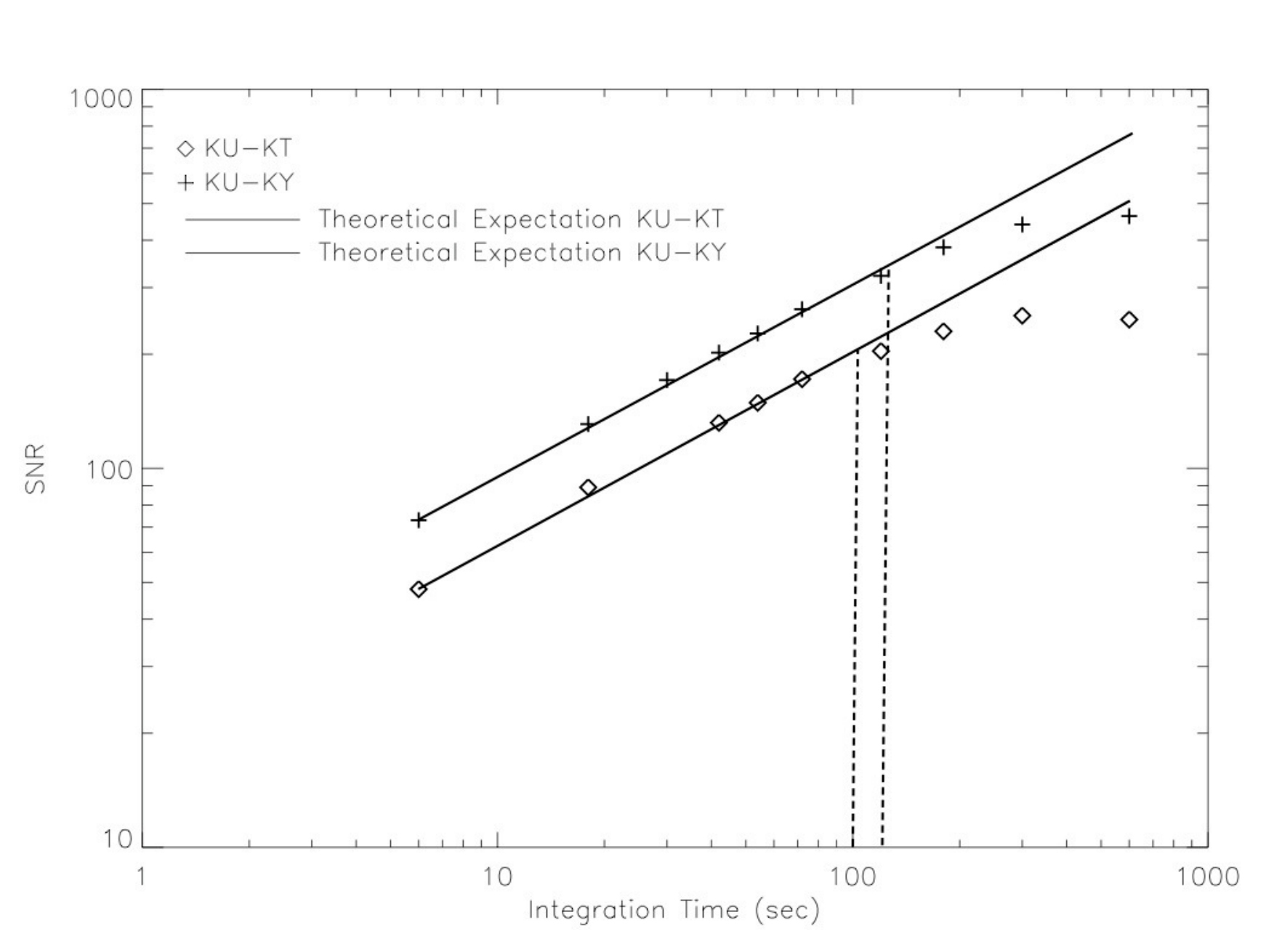}
\caption[The SNR level of 3C273 versus integration time for the KU-KT baseline (diamond sign) and the KU-KY (cross sign). The integration time corresponds to the different solution intervals. The shortest solution interval is at 0.1 min, while the longest one is set to be at $\sim$ 10 min, length of the entire scan. The KU telescope was used as a reference antenna. The theoretical expectation for each antenna is also over-plotted with solid black lines for each baseline in order to derive the relative coherence time. Additionally, a linear fitting is shown for both baselines (red color solid lines).]
{The SNR level of 3C273 versus integration time for the KU-KT baseline (diamond sign) and the KU-KY (cross sign). The integration time corresponds to the different solution intervals. The shortest solution interval is at 0.1 min, while the longest one is set to be at $\sim$ 10 min, length of the entire scan. The KU telescope was used as a reference antenna. The theoretical expectation for each antenna is also over-plotted with solid black lines for each baseline in order to derive the relative coherence time. The SNR of KU-KT baseline is deviated from the theoretical expectant at $\sim$ 120 sec and it becomes flat at 180 sec. The SNR of KU-KY baseline shows a deviation at $\sim$ 120 sec and is converged at 300 sec.}
\label{coherence}
\end{figure}

Figure \ref{coherence} shows the SNR of 3C273 as a function of integration time and describes the point where the visibility phase starts to lose its coherence. The SNR has been calculated for all baselines (KU-KT, KU-KY) using the Ulsan telescope as reference antenna.
The integration time is described in the horizontal axis corresponds to the different solution intervals. The shortest solution interval is 0.1 min, while the longest one was set to $\simeq$ 10 min, the entire scan length. The theoretical expectant value of the SNR for each baseline is also over-plotted with black solid lines. 
For the KU-KT baseline the SNR increases within 300 sec and after that time it becomes flat. In this case longer integration time than 300 sec does not provide a better SNR. Note that the observed SNR starts to deviate from the theoretical expectation in $\sim$ 100 sec. %For the case of KU-KY baseline the SNR starts to be converged at 180 sec and is deviated at $\sim$ 120 sec. 
For the case of KU-KY baseline, the SNR starts to deviate at $\sim$ 120 sec and begins to converge at $\sim$ 300 sec. The point where the prediction starts deviating from the actual measurement (vertical dashed lines) gives the idea where the radio signal begins to be affected by coherence loss.

\section{Summary}\label{sum}
 In this work, we have presented the results of eight radio sources in the Virgo cluster area from 22 GHz KVN VLBI observations. Particularly, the applications of phase referencing and its detection capability with KVN have been discussed for the first time. 
 
 Our targets provide an excellent sample to test the phase referencing capability of the KVN, according to the wide range of separation angles between the target and the calibrator and their estimated fluxes range from tens mJy to a few Jy. 

 %Practically the detection of the weak sources is not limited.
 Although the integration time can be extended as much as possible by PR, the detection is clearly limited due to the thermal noise which differs with the atmospheric conditions. 
However, PR can be affected by several factors. 
Beside the atmospheric conditions and the large separation angles the elevation of the sources affects the SNR, especially at low elevation where the propagation medium becomes longer and the correction with the atmospheric model is less accurate can harm the clear detection. 

Despite the fact of the non-detections, we successfully detected phase visibilities and we obtained the 22 GHz fluxes of three targets. The flux measurements with PR do not show any difference in comparison to the self-calibrated results, corroborating the reliable outcome of the KVN's PR. 

Additionally,  we have found the coherence time scale where the measured SNR starts to deviate from the theoretical expectation. The SNR increases as a function of the integration time but due to the atmospheric fluctuations the signal suffers from coherence loss. When the observations are affected by coherence loss, the delay measurements can potentially be less accurate (large delay errors). Hence the calibration of the atmospheric effects becomes hard and the detection of weak sources is challenging.
 
The sources that do not show a clear evidence of detection need to be carefully examined. In order to acquire more accurate measurements of their emission we plan a future test for the phasing referencing
with KVN and VLBI Exploration of Radio Astrometry (VERA) Array (KaVA).
 %technique combining both (KaVA) KVN observations with VLBI Exploration of Radio Astrometry (VERA). 
These future observations will evaluate not only phase referencing observations but  it will also give technical information such as the practical slewing / settling time for various separation angles, practical scheduling / operational issues and detection sensitivity.

%\acknowledgments
\section*{Acknowledgments}
We thank the anonymous referee for the useful comments that helped to improve the manuscript. We are grateful to all staff at the KVN who helped to operate the array and to correlate the data. The KVN is a facility operated by the Korea Astronomy and Space Science Institute. This paper is a result of the collaborative project between Korea Astronomy and Space Science Institute and Yonsei University through DRC program of Korea Research Council of Fundamental Science and Technology (DRC-12-2-KASI). This work has been also supported by the SRC program of the National Research Foundation of Korea and Science Fellowship of POSCO TJ Park Foundation.

%%%%%%%%%%%%%%%%%%%%%%%%%%%%%%%%%%%%%%%%%%%%%%%%%%%%%%%%%%%%%%%%%%%%%%%%%%%%%
%% Appendices
% \appendix

%\newpage 
%\appendix
%\section{Appendices}
% 


\section*{References}

\begin{thebibliography}{}

\bibitem[Alef(1988)]{1988IAUS..129..523A}
{Alef}, W. 1988, in IAU Symposium, Vol. 129, The Impact of VLBI on Astrophysics
  and Geophysics, ed. M.~J. {Reid} \& J.~M. {Moran}, 523
  
\bibitem[Asaki et al.(1998)]{1998RaSc...33.1297A}
{Asaki}, Y. and {Shibata}, K.~M. and {Kawabe}, R. and {Roh}, D.-G. and 
{Saito}, M. and {Morita}, K.-I. and {Sasao}, T. Phase compensation experiments with the paired antennas method 2. Millimeter-wave fringe correction using centimeter-wave reference, 33, 1297

\bibitem[Asaki et al.(1996)]{1996RaSc...31.1615A}
{Asaki}, Y. and {Saito}, M. and {Kawabe}, R. and {Morita}, K.-I. and 
	{Sasao}, T.  Phase compensation experiments with the paired antennas method, 31, 1615

\bibitem[Askne, J. \& Nordius, H. (1987)]{askne}
{Askne}, J. \& {Nordius}, H. (1987). Estimation of tropospheric delay for microwaves from surface weather data. Radio Science, Vol.22, No.3, pp. 379-386, ISSN 0048-6604  

\bibitem[Beasley et al.(1995)]{1995ASPC...82..327B}
{Beasley}, A.~J. and {Conway}, J.~E. 1995, in Astronomical 
Society of the Pacific Conference Series, Vol~82, 
Very Long Baseline Interferometry and the VLBA, ed. {Zensus}, J.~A. and {Diamond}, P.~J. and {Napier}, P.~J., 327

\bibitem[Becker et al.(1994)]{1994ASPC...61..165B}
{Becker}, R.~H., {White}, R.~L., \& {Helfand}, D.~J. 1994, in Astronomical
  Society of the Pacific Conference Series, Vol.~61, Astronomical Data Analysis
  Software and Systems III, ed. D.~R. {Crabtree}, R.~J. {Hanisch}, \&
  J.~{Barnes}, 165

\bibitem[Best et al.(2005)]{2005MNRAS.362...25B}
{Best}, P.~N., {Kauffmann}, G., {Heckman}, T.~M., {et~al.} 2005, \mnras, 362,
  25

\bibitem[Bower et al.(2000)]{2000ApJ...534..189B}
{Bower}, G.~A., {Green}, R.~F., {Quillen}, A.~C., {et~al.} 2000, \apj, 534, 189

\bibitem[Capetti et al.(2009)]{2009AJ....138.1990C}
{Capetti}, A., {Kharb}, P., {Axon}, D.~J., {Merritt}, D., \& {Baldi}, R.~D.
  2009, \aj, 138, 1990

\bibitem[Crane(1976)]{crane}
{Crane}, R.~K. 1976 in Astrophysics. Part B: Radio Telescopes, 186-200


\bibitem[Croton et al.(2006)]{2006MNRAS.365...11C}
{Croton}, D.~J., {Springel}, V., {White}, S.~D.~M., {et~al.} 2006, \mnras, 365,
  11

\bibitem[Davies \& Birkinshaw(1988)]{1988ApJS...68..409D}
{Davies}, R.~L. \& {Birkinshaw}, M. 1988, \apjs, 68, 409

\bibitem[Davis et al.(1985)]{Davis}
{Davis}, J. L., T. A. {Herring}, and I. I. {Shapiro}, A. E. E. {Rogers}, and G. 
{Elgered}, Geodesy by radio interferometry: Effects of atmospheric 
modelling errors on estimates of baseline length, 
Radio Sci., 20, 1593-1607, 1985 

\bibitem[Deller et al.(2007)]{Deller}
{Deller}, A.~T. and {Tingay}, S.~J. and {Bailes}, M. and {West}, C., 2007, in \pasp 119

\bibitem[Doi et al.(2006)]{2006PASJ...58..777D}
{Doi}, A. and {Fujisawa}, K. and {Habe}, A. and {Honma}, M. and 
	{Kawaguchi}, N. and {Kobayashi}, H. and {Murata}, Y. and {Omodaka}, T. and 
	{Sudou}, H. and {Takaba}, H., 2006, Bigradient Phase Referencing, \pasj, 58

\bibitem[Eracleous et al.(2010)]{2010ApJS..187..135E}
{Eracleous}, M., {Hwang}, J.~A., \& {Flohic}, H.~M.~L.~G. 2010, \apjs, 187, 135

\bibitem[Finoguenov et al.(2008)]{2008ApJ...686..911F}
{Finoguenov}, A., {Ruszkowski}, M., {Jones}, C., {et~al.} 2008, \apj, 686, 911

\bibitem[Fomalont(1981)]{1981NRAON...3....3F}
{Fomalont}, E. 1981, National Radio Astronomy Observatory Newsletter, 3, 3

\bibitem[Giacintucci et al.(2011)]{2011ApJ...732...95G}
{Giacintucci}, S., {O'Sullivan}, E., {Vrtilek}, J., {et~al.} 2011, \apj, 732,
  95

\bibitem[Gilmour et al.(2007)]{2007MNRAS.380.1467G}
{Gilmour}, R., {Gray}, M.~E., {Almaini}, O., {et~al.} 2007, \mnras, 380, 1467

\bibitem[Honma et al.(2007)]{2003PASJ...55L..57H}
{Honma}, M. and {Fujii}, T. and {Hirota}, T. and {Horiai}, K. and 
	{Iwadate}, K. and {Jike}, T. and {Kameya}, O. and {Kamohara}, R. and 
	{Kan-Ya}, Y. and {Kawaguchi}, N. and {Kobayashi}, H. and {Kuji}, S. and 
	{Kurayama}, T. and {Manabe}, S. and {Miyaji}, T. and {Nakashima}, K. and 
	{Omodaka}, T. and {Oyama}, T. and {Sakai}, S. and {Sakakibara}, S.-I. and 
	{Sato}, K. and {Sasao}, T. and {Shibata}, K.~M. and {Shimizu}, R. and 
	{Suda}, H. and {Tamura}, Y. and {Ujihara}, H. and {Yoshimura}, A., 2003, First Fringe Detection with VERA's Dual-Beam System and Its Phase-Referencing Capability, \pasj, 55

\bibitem[Jones \& Wehrle(1997)]{1997AAS...19110403J}
{Jones}, D.~L. \& {Wehrle}, A.~E. 1997, in Bulletin of the American
  Astronomical Society, Vol.~29, American Astronomical Society Meeting
  Abstracts, \#104.03
  
\bibitem[Jones et al.(1999)]{1999AAS...194.5018J}
{Jones}, D.~L., {Wehrle}, A.~E., \& {Piner}, B.~G. 1999, in Bulletin of the
  American Astronomical Society, Vol.~31, American Astronomical Society Meeting
  Abstracts \#194, 899
%

\bibitem[Jung et al.(2011)]{2011PASJ...63..717J}
{Jung}, T. and {Sohn}, B.~W. and {Kobayashi}, H. and {Sasao}, T. and 
	{Hirota}, T. and {Kameya}, O. and {Choi}, Y.~K. and {Chung}, H.~S., 2011, Erratum: First Simultaneous Dual-Frequency Phase Referencing VLBI Observation with VERA, \pasj, 63
	
	
\bibitem[Kim et al.(2004)]{2004evn..conf..281K}
{Kim}, H.-G., {Han}, S.-T., {Sohn}, B.~W., {et~al.} 2004, in European VLBI
  Network on New Developments in VLBI Science and Technology, ed.
  R.~{Bachiller}, F.~{Colomer}, J.-F. {Desmurs}, \& P.~{de Vicente}, 281--284

\bibitem[Kormendy et al.(2009)]{2009ApJS..182..216K}
{Kormendy}, J., {Fisher}, D.~B., {Cornell}, M.~E., \& {Bender}, R. 2009, \apjs,
  182, 216

\bibitem[Koulouridis \& Plionis(2010)]{2010ApJ...714L.181K}
{Koulouridis}, E. \& {Plionis}, M. 2010, \apjl, 714, L181

\bibitem[Lee et al.(2011)]{2011PASP..123.1398L}
{Lee}, S.-S., {Byun}, D.-Y., {Oh}, C.~S., {et~al.} 2011, \pasp, 123, 1398

\bibitem[Ly et al.(2004)]{2004AJ....127..119L}
{Ly}, C., {Walker}, R.~C., \& {Wrobel}, J.~M. 2004, \aj, 127, 119

\bibitem[Marvel \& Woody(1998)]{1998AAS...192.8103M}
{Marvel}, K.~B. and {Woody}, D.~P. Phase Correction at Millimeter Wavelengths Using Observations of Atmospheric Water Vapor at 22 GHz, 30, 1154

\bibitem[Mei et al.(2007)]{2007ApJ...655..144M}
{Mei}, S., {Blakeslee}, J.~P., {C{\^o}t{\'e}}, P., {et~al.} 2007, \apj, 655,
  144
  
\bibitem[Miley(1980)]{1980ARA&A..18..165M}
{Miley}, G. 1980, \araa, 18, 165

\bibitem[Middelberg et al.(2005)]{2005A&A...433..897M}
{Middelberg}, E. and {Roy}, A.~L. and {Walker}, R.~C. and {Falcke}, H. , VLBI observations of weak sources using fast frequency switching, \aap, 433

\bibitem[Moran \& Rosen(1981)]{1981RaSc...16..235M}
{Moran}, J.~M. and {Rosen}, B.~R. Estimation of the propagation delay through the troposphere from microwave radiometer data, 16, 235-244

\bibitem[Murphy et al.(2009)]{2009ApJ...694.1435M}
{Murphy}, E.~J., {Kenney}, J.~D.~P., {Helou}, G., {Chung}, A., \& {Howell},
  J.~H. 2009, \apj, 694, 1435
  
\bibitem[Nagar et al.(2005)]{2005A&A...435..521N}
{Nagar}, N.~M. and {Falcke}, H. and {Wilson}, A.~S.,
  \aap, 435, 521-543

\bibitem[Parma et al.(2002)]{2002NewAR..46..313P}
{Parma}, P., {Murgia}, M., {de Ruiter}, H.~R., \& {Fanti}, R. 2002, \nar, 46,
  313
  
\bibitem[Porcas et al.(2003)]{2003ASPC..306...39P}
{Porcas}, R.~W. and {Alef}, W. and {Rioja}, M.-J. and {Desmurs}, J.-F. and 
	{Gurvits}, L.~I. and {Schilizzi}, R.~T., Multi-view VLBI with Arrays in Cluster-Cluster Mode, 306, 29
  
\bibitem[Rioja et al.(2002)]{2002evn..conf...57R}
{Rioja}, M.~J. and {Porcas}, R.~W. and {Desmurs}, J.-F. and 
	{Alef}, W. and {Gurvits}, L.~I. and {Schilizzi}, R.~T. , VLBI observations in Cluster-Cluster mode at 1.6 GHz, 2002, 57
  
 \bibitem[Roy et al.(2004)]{2004evn..conf..265R}
{Roy}, A.~L. and {Teuber}, U. and {Keller}, R. , The Water Vapour Radiometer at Effelsberg, 2004, 265-270 
  
\bibitem[Saastamoinen et al.(1972)]{Saastamoinen}
{Saastamoinen}, J. Atmospheric Correction for Troposphere and Stratosphere in Radio Ranging of Satellites, The Use of Artificial Satellites for Geodesy, 15, 247

\bibitem[Sambruna et al.(2003)]{2003ApJ...586L..37S}
{Sambruna}, R.~M. and {Gliozzi}, M. and {Eracleous}, M. and  {Brandt}, W.~N. and {Mushotzky}, R., 2003
\apjl,  586

\bibitem[Sasao et al.(2003)]{2003ASPC..306...53S}
{Sasao}, T., J. Technologies for mm and Sub-mm VLBI: Multi-frequency Feed as a Tool for mm-Wave VLBI, 306, 53

\bibitem[Shabala \& Alexander(2009)]{2009ApJ...699..525S}
{Shabala}, S. \& {Alexander}, P. 2009, \apj, 699, 525

\bibitem[Shabala et al.(2011)]{2011gafo.confP.159S}
{Shabala}, S., {Kaviraj}, S., \& {Silk}, J. 2011, in Galaxy Formation, 159P

\bibitem[Shabala et al.(2008)]{2008MNRAS.388..625S}
{Shabala}, S.~S., {Ash}, S., {Alexander}, P., \& {Riley}, J.~M. 2008, \mnras,
  388, 625

\bibitem[Shapiro et al.(1979)]{1979AJ.....84.1459S}
{Shapiro}, I.~I., {Wittels}, J.~J., {Counselman}, III, C.~C., {et~al.} 1979,
  \aj, 84, 1459

\bibitem[Shepherd(1997)]{1997ASPC..125...77S}
{Shepherd}, M.~C. 1997, in Astronomical Society of the Pacific Conference
  Series, Vol. 125, Astronomical Data Analysis Software and Systems VI, ed.
  G.~{Hunt} \& H.~{Payne}, 77
  
\bibitem[Shin et al.(2012)]{2012ApJ...745...13S}
{Shin}, M.-S., {Ostriker}, J.~P., \& {Ciotti}, L. 2012, \apj, 745, 13

\bibitem[Thompson, A.R., Moran, J.M., and Swenson, G.W.(1986)]{tms}
{Thompson}, A.R., {Moran}, J.M., and {Swenson}, G.W., Interferometry and Synthesis in Radio Astronomy (Wiley - Interscience, NY), 1986

\bibitem[Van der Laan \& Perola(1969)]{1969A&A.....3..468V}
{Van der Laan}, H. \& {Perola}, G.~C. 1969, \aap, 3, 468

\bibitem[Vollmer et al.(2005)]{2005astro.ph..7252V}
{Vollmer}, B. and {Braine}, J and {Combes}, F. and {Sofue}, Y. 2005 \aap

\bibitem[Wright et al.(1996)]{1996PASP..108..520W}
{Wright}, M.~C.~H. Atmospheric Phase Noise and Aperture Synthesis Imaging at Millimeter Wavelengths 1996, \pasp, 108, 520

\bibitem[Xilouris \& Papadakis(2002)]{2002A&A...387..441X}
{Xilouris}, E.~M. \& {Papadakis}, I.~E. 2002, \aap, 387, 441

\end{thebibliography}
\end{document}